\documentclass[aps,prb,reprint,superscriptaddress,floatfix,longbibliography]{revtex4-2}
\usepackage{amsmath,amssymb,bm,graphicx,xcolor}
\usepackage[colorlinks=true,citecolor=blue,linkcolor=blue,urlcolor=blue]{hyperref}

\newcommand{\dd}{\mathrm{d}}

\begin{document}

\title{Analytical treatment of reentrant localization via Lyapunov branch structure}
\author{Taylan Yildiz}
\author{B.~Tanatar}
\affiliation{Department of Physics, Bilkent University, 06800 Ankara, T\"urkiye}
\date{\today}

\begin{abstract}
In reentrant localization, states become localized under a growing quasiperiodic potential partly become extended again before localizing once more. So far this has been demonstrated only numerically. Here we prove it in a chain of weakly coupled dimers in which we vary only the overall strength of a fixed onsite potential. As the strength increases, the lower orbital of each dimer shifts its weight between the two sites. At a particular strength, the quasiperiodic potential it feels cancels, while the hopping between dimers stays finite. When the energy scales are well separated, we prove that, over a finite range of potential strengths, all states in a central part of the lower band, one third of its states, have a vanishing Lyapunov exponent that is stable against small changes of phase, the certification of extended states. On both sides of this range, all states are localized. Numerical calculations confirm that the recovery covers the whole lower band, while the upper band stays localized.
\end{abstract}

\maketitle

\section{Introduction}
\label{sec:intro}

Anderson localization shows how in a disordered medium, interference can confine
waves and suppress diffusion
\cite{1}. In one dimension, uncorrelated
onsite disorder produces localization even at weak disorder
\cite{2}. Quasiperiodic potentials provide a
different setting: their deterministic  correlations
allow localization transitions at finite modulation
strength. The Aubry-Andr\'e (AA) model \cite{3,4,5} is the standard example,
with a self-dual transition criteria, the potential amplitude
equals twice the hopping \cite{6}. Localization in such models is most often
characterized through the Lyapunov exponent, or equivalently the
localization length, and through participation ratios and the multifractal
structure of eigenstates~\cite{66,67}.
Beyond these probes, localization in AA-type models has also been
studied through bulk polarization and geometric Binder
cumulants~\cite{59,60}.

Extensions of the AA model reveal a richer spectral
structure. Longer-range hopping and changes in the onsite
potential can produce mobility edges, separating localized
and extended states at different energies
\cite{7,8,11,13,14,15,16,17,18}. Non-Hermitian models extend
these questions to asymmetric hopping and complex onsite
potentials, where localization is connected to changes in
the complex spectrum and, in some models, symmetry
breaking \cite{19,20,21,22,23,24,25,26,27,28}. The AA model and its generalizations have also been
studied on ladders \cite{9,57},
two-dimensional lattices \cite{58,10}, diamond chains
\cite{56}, and helical structures
\cite{54,55}. These examples
show that localization depends on both the strength and
the structure of the modulation.

Quasiperiodic localization has been observed with
ultracold atoms in optical lattices \cite{29}
and light in photonic waveguide arrays \cite{30}.
In atomic systems, interactions can be tuned while keeping
the spatial potential fixed. This connects quasiperiodic
localization to many-body localization, where an isolated
interacting system can retain local memory and fail to
thermalize \cite{31,32}. Experiments with
interacting fermions in a quasiperiodic optical lattice
have observed a density imbalance that persists over the
accessible times \cite{33}. Related studies
of interacting localization have used trapped-ion
simulators \cite{34} and superconducting circuits
\cite{35}.

At the single-particle level, the standard AA model gives
a monotonic response to increasing modulation: its states
become localized beyond a single threshold and remain
localized as the potential grows. Reentrant localization
changes this sequence. A further increase in an onsite
control can restore spatial extent in part of the spectrum,
followed by localization again at larger coupling.
Such behavior has been reported in dimerized
quasiperiodic chains \cite{36,37,51,52,53}, models with an additional
staggered potential \cite{38,48,49}, lattices with longer-range
hopping \cite{39,40,50}, helical systems \cite{41,42}, and
non-Hermitian quasiperiodic lattices \cite{43,44,45,46}.

Here we study this question in a nearest-neighbor chain
of weakly coupled dimers. All hoppings and the spatial
onsite profile remain fixed while one overall potential
amplitude is varied. This amplitude changes both the local
detuning and the quasiperiodic modulation. The resulting
redistribution of dimer weights cancels the first-order
potential in the lower band while leaving its hopping
finite. The upper band has no such cancellation, so the
recovery is band selective. Wang et al. \cite{61}
identified orbital-weight cancellation numerically in a
trimer ladder; here we establish its consequences for the full dimer chain, beyond the projection. We place the weak hopping between the
residual modulation scale near cancellation and the larger
modulation away from it.

Near cancellation, the exact transfer matrix remains
uniformly close to a constant rotation throughout a fixed
complex strip of phases. This bound limits the complexified
Lyapunov exponent on the strip. Since its slopes are
quantized by Avila's global theory~\cite{62}, the
exponent must be flat near the real axis, and a flat
positive exponent is excluded on the spectrum. For
sufficiently small modulation parameter $\eta$, every
spectral energy in a central lower-band window is therefore
subcritical, with a finite zero-exponent plateau,
throughout a recovery interval of couplings of width of order $a/\eta$.
Continuity of the rotation number shows that this window holds one third
of the lower-band states.   
The exponents are uniformly positive at comparison couplings
before and after recovery. 

The branches of the complexified Lyapunov exponent also
provide a diagnostic at finite parameters. A signed branch
intercept summarizes the loss and recovery of finite
zero-exponent plateaus across the spectrum and remains
informative when the physical exponent vanishes.
Spectrum-wide participation maps and plateau scans show
that the recovery covers the entire lower band and is
absent in the upper band. The numerically determined
window edges coincide with the crossings predicted by the
first-order projection, and the window width follows the
predicted $a/\eta$ scaling as $\eta$ decreases.

Section~\ref{sec:model} introduces the model and its dimer
projection. Section~\ref{sec:branches} develops the
Lyapunov branch diagnostic, and
Sec.~\ref{sec:controlled} establishes the controlled
reentrant regime. Section~\ref{sec:numerics} presents
the numerical results, and Sec.~\ref{sec:discussion}
discusses their implications and scope. Technical
derivations are collected in the appendices.

\section{Model and dimer mechanism}
\label{sec:model}

We consider an infinite chain with Hamiltonian
\begin{align}
 H={}&\sum_m\bigl[b\,c_{A,m}^{\dagger}c_{B,m}
       +a\,c_{B,m}^{\dagger}c_{A,m+1}+\mathrm{H.c.}\bigr]
       \nonumber\\
 &+\sum_m\bigl[\varepsilon_{A,m}(\lambda)n_{A,m}
                 +\varepsilon_{B,m}(\lambda)n_{B,m}\bigr],
 \label{eq:H}
\end{align}
with $a,b>0$ and $a\ll b$. The strong bond joins $A_m$ and $B_m$ while
the weak bond joins neighboring cells. The onsite terms are
\begin{align}
 \varepsilon_{A,m}(\lambda)&=\lambda[1+3\eta\cos\theta_m],\nonumber\\
 \varepsilon_{B,m}(\lambda)&=-\lambda[1+\eta\cos\theta_m],\quad
 \theta_m=2\pi\beta m+\phi.
 \label{eq:onsite}
\end{align}
The frequency $\beta$ is irrational to ensure quasiperiodicity. The parameters $a$, $b$,
$\eta>0$, and $\beta$ remain fixed while $\lambda\ge0$ is varied. Thus
$H=H_{\rm hop}+\lambda F$ with one fixed onsite profile $F$.

The quasiperiodic terms have opposite signs, unequal magnitudes, and a
common phase within each cell. This choice is central to solve analytically. It differs from a cosine sampled at every original
site, where adjacent sites have a phase difference.

\subsection{Projected potential and hopping}
\label{sec:projection}

We first solve the strong dimers exactly, keeping the detuning
$\lambda$, and then treat the quasiperiodic modulation and weak bonds
as perturbations. For $a,b>0$, the isolated dimer has
\begin{align}
 h_0(\lambda)&=
 \begin{pmatrix}\lambda&b\\ b&-\lambda\end{pmatrix}, \quad
 E_\pm(\lambda)=\pm\sqrt{b^2+\lambda^2}.
 \label{eq:dimer}
\end{align}
Its level separation is at least $2b$. The lower state has site
probabilities
\begin{equation}
 w_A=\frac12\left(1-\frac{\lambda}{\sqrt{b^2+\lambda^2}}\right),\quad
 w_B=\frac12\left(1+\frac{\lambda}{\sqrt{b^2+\lambda^2}}\right),
 \label{eq:weights}
\end{equation}
and can be written as
\begin{equation}
 |m,-\rangle=\sqrt{w_A}\,|A,m\rangle
             -\sqrt{w_B}\,|B,m\rangle.
\end{equation}
As $\lambda$ grows, its weight shifts toward the lower-energy
site $B$.

All lower dimer states have the same unperturbed energy.
We therefore allow the perturbations to mix these states and
calculate their diagonal and nearest-neighbor matrix elements.
Coupling through the upper dimer states contributes at higher order.

The quasiperiodic onsite perturbation is
\begin{equation}
 \delta h_m=\eta\lambda\cos\theta_m
 \begin{pmatrix}3&0\\0&-1\end{pmatrix},
\end{equation}
Its first-order contribution to the local energy is
\begin{align}
 V_{-,m}
 &=\langle m,-|\delta h_m|m,-\rangle\nonumber\\
 &=\eta\lambda(3w_A-w_B)\cos\theta_m\nonumber\\
 &=\eta\lambda\left(1-\frac{2\lambda}
 {\sqrt{b^2+\lambda^2}}\right)\cos\theta_m.
 \label{eq:vminus}
\end{align}
Writing $V_{-,m}=v_-(\lambda)\cos\theta_m$, the modulation amplitude
is thus the probability-weighted sum of the two opposite-sign
onsite potentials. The contributions cancel at
\begin{equation}
 \lambda_*=\frac{b}{\sqrt3},\qquad
 w_A(\lambda_*)=\frac14,\qquad w_B(\lambda_*)=\frac34.
 \label{eq:gstar}
\end{equation}
The smaller probability on $A$ exactly compensates its larger
modulation coefficient.

The weak bond between adjacent dimers,
\begin{equation}
 H_{a,m}=a|B,m\rangle\langle A,m+1|+\mathrm{H.c.},
\end{equation}
gives the first-order hopping matrix element
\begin{align}
 t_{\rm eff}(\lambda)
 &=\langle m,-|H_{a,m}|m+1,-\rangle\nonumber\\
 &=a(-\sqrt{w_B})\sqrt{w_A}
 =-a\sqrt{w_Aw_B}.
\end{align}
Since
\begin{equation}
 w_Aw_B
 =\frac14\left(1-\frac{\lambda^2}{b^2+\lambda^2}\right)
 =\frac{b^2}{4(b^2+\lambda^2)},
\end{equation}
its magnitude is
\begin{align}
 |t_{\rm eff}(\lambda)|
 =\frac{ab}{2\sqrt{b^2+\lambda^2}},\quad\text{and}\quad
 |t_{\rm eff}(\lambda_*)|&=\frac{\sqrt3}{4}a.
 \label{eq:teff}
\end{align}
The effective hopping therefore remains finite when the first-order
modulation cancels.

\begin{figure}[h]
    \centering
    \includegraphics[width=0.9\linewidth]{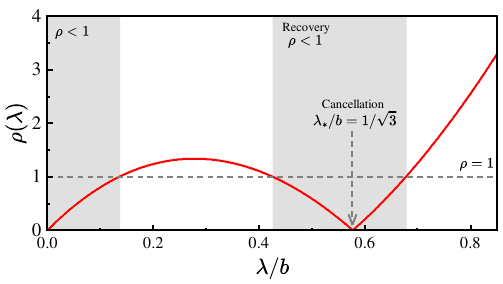}
    \caption{Projected modulation-to-hopping ratio for
    $a/b=10^{-3}$ and $\eta=10^{-2}$. Shading marks
    $\rho<1$; the arrow indicates first-order cancellation.}
    \label{fig:2}
\end{figure}

Writing the state in this description as
$|\Psi\rangle=\sum_m\chi_m|m,-\rangle$,
first-order perturbation gives
\begin{align}
 \left(E+\sqrt{b^2+\lambda^2}\right)\chi_m
 &=-|t_{\rm eff}|(\chi_{m+1}+\chi_{m-1})\nonumber\\
 &\quad+v_-(\lambda)\cos\theta_m\,\chi_m.
\end{align}
The change of amplitudes $\chi_m=(-1)^m u_m$ gives
$\chi_{m\pm1}=-(-1)^m u_{m\pm1}$ and reverses the hopping sign.
The effective equation becomes
\begin{align}
 \left(E+\sqrt{b^2+\lambda^2}\right)u_m
 &=|t_{\rm eff}|(u_{m+1}+u_{m-1})\nonumber\\
 &\quad+v_-(\lambda)\cos\theta_m\,u_m.
 \label{eq:effective}
\end{align}
To this order, the lower band is therefore described by an AA chain
with modulation-hopping ratio
\begin{align}
 \rho(\lambda)
 &=\frac{|v_-(\lambda)|}{2|t_{\rm eff}(\lambda)|}
 =\frac{\eta\lambda
 \left|\sqrt{b^2+\lambda^2}-2\lambda\right|}{ab}.
 \label{eq:Xi}
\end{align}
On its spectrum, this effective AA model has zero Lyapunov exponent
for $\rho\le1$ and a positive exponent for $\rho>1$
\cite{3}.
For $a\ll\eta b$, the ratio can exceed one before $\lambda_*$,
vanishes at $\lambda_*$, and grows again  (Fig. \ref{fig:2}). 

For the upper dimer state, the site probabilities are interchanged,
giving the first-order modulation amplitude
\begin{equation}
 v_+(\lambda)=\eta\lambda
 \left(1+\frac{2\lambda}{\sqrt{b^2+\lambda^2}}\right).
\end{equation}
It has no cancellation at positive $\lambda$, so the first-order
recovery is band selective.

The projection predicts the physical route to recovery, but its AA
criterion is exact only for Eq.~\eqref{eq:effective}. The original
chain retains interband processes and higher-order modulation. Their
control is developed in Sec.~\ref{sec:controlled}.

\section{Lyapunov branch structure}
\label{sec:branches}

\subsection{Transfer matrices and imaginary phase}
\label{sec:transfer}

Write a stationary wavefunction as $(x_m,y_m)$ on $(A_m,B_m)$.
The site equations are
\begin{align}
 (E-\varepsilon_{A,m})x_m&=b y_m+a y_{m-1},\nonumber\\
 (E-\varepsilon_{B,m})y_m&=b x_m+a x_{m+1}.
 \label{eq:site}
\end{align}
With $P=E-\varepsilon_A(\theta)$ and
$Q=E-\varepsilon_B(\theta)$, the exact cell transfer matrix is
\begin{align}
 \begin{pmatrix}x_{m+1}\\y_m\end{pmatrix}
 &=M_E(\theta_m)\begin{pmatrix}x_m\\y_{m-1}\end{pmatrix},\nonumber\\
 M_E(\theta)&=\frac1{ab}
 \begin{pmatrix}PQ-b^2&-aQ\\aP&-a^2\end{pmatrix}.
 \label{eq:M}
\end{align}
Its determinant is one. The chain samples this one matrix-valued
function along $\theta\mapsto\theta+2\pi\beta$; this transfer
evolution is the cocycle. Complexify the phase as $\phi\to\phi+ih$.
Let
\begin{equation}
    \mathcal M_N(E,\phi+ih)=M_E(\phi+(N-1)2\pi\beta+ih)
\cdots M_E(\phi+ih)
\end{equation}
and define
\begin{equation}
 L(E,h)=\lim_{N\to\infty}\frac1N\frac1{2\pi}
       \int_0^{2\pi}\ln\|\mathcal M_N(E,\phi+ih)\|\,\dd\phi.
 \label{eq:L}
\end{equation}
This exponent is per cell. The physical exponent per original site reads as
\begin{equation}
 \gamma(E,\lambda)=\frac12 L(E,0).
 \label{eq:gamma}
\end{equation}

The imaginary phase is an analytical probe of the real chain.
The identity
$\cos(\theta+ih)=(e^{-h+i\theta}+e^{h-i\theta})/2$
shows how it amplifies one component. $L(E,h)$ is even and convex, with integer slopes for this
analytic determinant-one cocycle \cite{62}. 

For our model and $h\ge0$, the complexified Lyapunov exponent is
piecewise linear with allowed slopes zero, one, and two.
The upper bound follows from the Fourier degree of the cell transfer
matrix. We characterize the slope-one and slope-two branches by their
intercepts,
\begin{align}
 c_1(E,\lambda)&=\min_{h\ge0}[L(E,h)-h],
 \label{eq:c1}\\
 c_2(\lambda)&=\ln\frac{3\eta^2\lambda^2}{4ab}
 =2\ln\frac{\lambda}{\lambda_H},
 \label{eq:c2}
\end{align}
where
\begin{equation}
 \lambda_H=\frac{2\sqrt{ab}}{\sqrt3\,\eta}.
 \label{eq:lambdaH}
\end{equation}
For $\lambda>\lambda_H$, the inequality
$L(E,0)\ge c_2(\lambda)$ forces every spectral exponent to be positive.
Thus $\lambda_H$ is an upper bound on the survival of zero exponents,
rather than an exact transition point. The slope restriction, $c_1$ construction, the intercept
formulas, and the resulting bound are detailed in
Appendix~\ref{app:branches}.

\subsection{Diagnostic across the transition}
\label{sec:signed}

For an energy in a band, a flat segment can occur only at height zero. A
positive flat segment would imply uniform hyperbolicity and place the
energy outside the spectrum.

We have the complexified Lyapunov exponent as
\begin{equation}
 L(E,h)=\max\{0,c_1(E,\lambda)+h,c_2(\lambda)+2h\}
 \label{eq:envelope}
\end{equation}

For a zero-exponent energy, let $h_c$ be the largest $h\ge0$ for
which $L(E,h)=0$; set $h_c=0$ when the physical exponent is positive.
Eq.~\eqref{eq:envelope} gives
\begin{equation}
 c_1(E,\lambda)=
 \begin{cases}
  2\gamma(E,\lambda),&\gamma(E,\lambda)>0,\\
  -h_c(E,\lambda),&\gamma(E,\lambda)=0.
 \end{cases}
 \label{eq:meaning}
\end{equation}
Thus $c_1>0$ identifies a positive exponent, $c_1<0$ a subcritical
energy with a finite zero plateau, and $c_1=0$ a critical energy.
The plateau width measures stability under imaginary phase. To follow the chain as a whole, we define
\begin{align}
 C_1(\lambda)=\min_{E\in\sigma(H_\lambda)}
 c_1(&E,\lambda),\quad
 D(\lambda)=\min_{E\in\sigma(H_\lambda)}
 \gamma(E,\lambda),\nonumber\\
 W(\lambda)&=\max_{E\in\sigma(H_\lambda)}h_c(E,\lambda).
 \label{eq:global}
\end{align}
where $\sigma(H_\lambda)$ denotes the spectrum. They obey
\begin{equation}
 2D=\max\{0,C_1\},\qquad W=\max\{0,-C_1\}.
 \label{eq:DW}
\end{equation}
A positive $C_1$ means that every spectral energy has a positive
exponent. A negative value measures the widest subcritical plateau.
At $C_1=0$, critical energies can remain even though $W=0$.
Recovery of finite plateaus between positive-exponent regimes is
therefore the sequence $C_1>0$, $C_1<0$, $C_1>0$.

Below $\lambda_H$, the slope-two line limits every zero plateau to
\begin{equation}
 h_c(E,\lambda)\le h_{\max}(\lambda)
 =-\frac{c_2(\lambda)}2=\ln\frac{\lambda_H}{\lambda}.
 \label{eq:hmax}
\end{equation}
For $E\in\sigma(H_\lambda)$ and $0<\lambda<\lambda_H$,
Eq.~\eqref{eq:envelope} then gives a useful identity,
\begin{equation}
 c_1(E,\lambda)=L(E,h_{\max})-h_{\max}.
 \label{eq:oneheight}
\end{equation}

\section{ reentrant regime}
\label{sec:controlled}

\subsection{Controlled subcritical recovery}
\label{sec:recovery}

The first-order perturbation predicts cancellation of the
lower-cell modulation at $\lambda_*=b/\sqrt3$. We now estimate
the modulation that remains beyond first order and compare
it with the hopping between cells.

The isolated cell can be diagonalized exactly, giving the
lower energy
\begin{equation}
 e_-(\theta,\lambda)=\eta\lambda\cos\theta
 -\sqrt{b^2+\lambda^2(1+2\eta\cos\theta)^2}.
 \label{eq:exactlocal}
\end{equation}
At $\lambda_*$, the unperturbed lower-cell energy used in the
projection is
\[
 e_-^{(0)}(\lambda_*)
 =-\sqrt{b^2+\lambda_*^2}
 =-\frac{2b}{\sqrt3}.
\]
Expanding the exact energy around $\lambda=\lambda_*$
\begin{equation}
 e_-(\theta,\lambda_*)=e_-^{(0)}(\lambda_*)
 -\frac{\sqrt3}{4}\eta^2b\cos^2\theta+O(\eta^3b).
 \label{eq:residual}
\end{equation}
The first-order correction in $\eta$ vanishes, leaving a
correction of order $\eta^2b$. Using
$\cos^2\theta=[1+\cos(2\theta)]/2$, we separate this correction
into a uniform energy shift and a second-harmonic modulation.
The uniform shift changes every cell energy equally; the
second harmonic produces the remaining energy differences
between cells.

The exact bound
\begin{equation}
 -\frac23\eta^2b
 \le e_-(\theta,\lambda_*)-e_-^{(0)}(\lambda_*)\le0
 \label{eq:residualbound}
\end{equation}
holds for every $\theta$, including all higher-order corrections.
Thus the lower-cell energies lie within an interval of width
at most $2\eta^2b/3$.

Equation~\eqref{eq:residualbound} limits the spread of the
lower-cell energies to order $\eta^2b$. The hopping between
these states remains of order $a$, with leading value
$|t_{\rm eff}|=\sqrt3a/4$ at $\lambda_*$. If $\eta^2b\ll a$,
the energy differences between cells are small compared with
their coupling, allowing the lower-cell states to hybridize
strongly.

Away from cancellation, at couplings of order $b$, the
first-order modulation is instead of order $\eta b$.
Choosing $a\ll\eta b$ makes this modulation strong compared
with the hopping. We therefore place the hopping between
the two modulation scales
\[
\eta^2b
 \ll a\ll
\eta b
\]
This hierarchy provides the energy-scale separation needed
for recovery near $\lambda_*$ between regimes of stronger
modulation.

We enforce both inequalities by choosing
\begin{equation}
 a=b\eta^p,\qquad 1<p<2,
 \label{eq:scaling}
\end{equation}
with fixed $b>0$, $p$, and $\beta$, and taking $\eta$
sufficiently small. Indeed,
$\eta^2b/a=\eta^{2-p}\to0$ and
$a/(\eta b)=\eta^{p-1}\to0$.

To control propagation in the original chain, we introduce
$r=a/b$ and write
\[
 E=-\frac{2b}{\sqrt3}+ax.
\]
Thus $x$ measures energy relative to the unperturbed lower-cell
level in units of the weak hopping. At $\lambda_*$,  as $\eta\to0$ with $r=\eta^p$ and
fixed $x$, the exact
transfer matrix converges to a phase independent matrix. We denote this
limiting matrix by
\begin{equation}
 M_0(x)=
 \begin{pmatrix}
 -4x/\sqrt3&1/\sqrt3\\
 -\sqrt3&0
 \end{pmatrix}.
 \label{eq:M0}
\end{equation}
We restrict the scaled energy with $ |x|\le\sqrt3/4$ . Throughout this window,
\[
 |\operatorname{tr}M_0(x)|
 =\frac{4|x|}{\sqrt3}\le1,
\]
so the limiting matrices remain uniformly inside the
elliptic regime $|\operatorname{tr}M_0|<2$. Each can therefore
be transformed into a rotation by a change of basis.

Near cancellation, the exact transfer matrix
is a small quasiperiodic perturbation of this elliptic matrix.
The uniform matrix estimate also constrains the Lyapunov
branches. After a bounded constant change of basis, the
exact transfer matrix is close to a rotation throughout
a fixed complex strip. Its Lyapunov exponent is therefore
small throughout that strip. Since every nonzero slope
is at least one, a rising branch cannot begin close to
$h=0$ without exceeding this bound. For sufficiently
small $\eta$, the exponent must have a plateau around
the real axis.

At a spectral energy, this plateau must have zero height.
A positive exponent that is locally independent of $h$
would imply uniform hyperbolicity, which
is excluded on the spectrum \cite{63}.
Thus every spectral energy $E=-2b/\sqrt3+ax$ with
$|x|\le\sqrt3/4$ satisfies
\begin{equation}
 L(E,h)=0,\qquad |h|\le h_0,
 \label{eq:spectralrecovery}
\end{equation}
where $h_0>0$ can be chosen uniformly over this energy
window. All spectral energies in the window are
therefore subcritical.

The window is also occupied. Near cancellation, the rotation number of the
exact transfer matrix stays close to that of $M_0(x)$, which sweeps an
interval of length $1/6$ across the window. Through the integrated density
of states, the window therefore holds a fraction $1/3+O(\delta_\eta)$ (with $\delta_\eta$ defined in Eq. \ref{eq:stripbound}) of
the lower-band states, all of them subcritical (Appendix \ref{app:subcritical}).

Recovery persists over a finite range of $\lambda$.
Moving away from $\lambda_*$ restores modulation of order
$\eta|\lambda-\lambda_*|$, while the hopping remains of
order $a$. Their competition sets the natural recovery
scale $a/\eta$.

Appendix~\ref{app:interval} shows that there is a constant
$c_*>0$ such that, for sufficiently small $\eta$, at every coupling satisfying
\begin{equation}
 |\lambda-\lambda_*|<c_*\frac{a}{\eta}.
 \label{eq:rigorouswindow}
\end{equation}
the controlled window holds a fraction close to one third of the
lower-band states, and every spectral energy in it is subcritical. This bound guarantees recovery within the stated interval
without locating the transition points. The recovered
energies follow the lower dimer level
$-\sqrt{b^2+\lambda^2}$ as $\lambda$ varies.

Within the first-order AA description, the condition
$\rho<1$ gives the leading width estimate
\begin{equation}
 |\lambda-\lambda_*|\lesssim\frac{2a}{\sqrt3\,\eta}.
 \label{eq:projectedwindow}
\end{equation}
Its derivation is also given in Appendix~\ref{app:interval}.

\subsection{Positive exponents before and after recovery}
\label{sec:checkpoints}

At the earlier comparison point $\lambda_{\rm e}=b/4$,
the original transfer matrix gives a uniform bound over
the spectrum,
\begin{equation}
 \gamma(E,\lambda_{\rm e})\ge\frac12
 \left[\ln\frac{\eta b}{a}-C_0\right]>0,
 \label{eq:early}
\end{equation}
for sufficiently small $\eta$ under Eq.~\eqref{eq:scaling}.
The constant $C_0$ is independent of $\eta$.
Appendix~\ref{app:early} derives this estimate from transfer
growth at a fixed imaginary phase and uses the maximum slope
two to obtain the real-axis bound.

For the later comparison point, choose
$\lambda_{\rm l}=2\lambda_H=4\sqrt{ab}/(\sqrt3\,\eta)$.
The bound in Eq.~\eqref{eq:physicalbound} gives
\begin{equation}
 \gamma(E,\lambda_{\rm l})
 \ge\ln\frac{\lambda_{\rm l}}{\lambda_H}
 =\ln2
 \label{eq:late}
\end{equation}
at every spectral energy. Moreover,
\begin{equation}
 \frac{\lambda_{\rm l}}{\lambda_*}
 =4\eta^{p/2-1}\longrightarrow\infty.
 \label{eq:ordering}
\end{equation}
The guaranteed recovery width, measured in units of $b$,
is $2c_*a/(\eta b)=2c_*\eta^{p-1}\to0$.
Together with $\lambda_{\rm e}<\lambda_*$ and
Eq.~\eqref{eq:ordering}, this places the comparison points
on opposite sides of the recovered interval for sufficiently
small $\eta$.

The minimum physical exponent therefore satisfies
\begin{align}
 &D(\lambda_{\rm e})>0,\qquad
 D(\lambda_{\rm l})\ge\ln2,\nonumber\\
 &D(\lambda)=0\quad\text{for}\quad
 |\lambda-\lambda_*|<c_*a/\eta.
 \label{eq:reentrantD}
\end{align}
The recovered energies have finite zero-exponent plateaus,
giving the branch signs
\begin{equation}
 C_1(\lambda_{\rm e})>0,\qquad
 C_1(\lambda_*)<0,\qquad
 C_1(\lambda_{\rm l})>0.
 \label{eq:reentrantC}
\end{equation}
At $\lambda=0$, the chain is periodic and its spectral
bands have zero exponent. The fixed-profile sweep therefore
exhibits an initial closing, a recovery, and a later closing
in the exponent-based sense.

The analytical bounds establish reentrance for sufficiently
small $\eta$ under Eq.~\eqref{eq:scaling}. The numerical
parameters used below follow the same scaling with
$p=3/2$, but their finite value of $\eta$ is not certified
by our conservative estimates. 

\section{Numerical results}
\label{sec:numerics}

Unless stated otherwise, we use $a/b=10^{-3}$,
$\eta=10^{-2}$, and the irrational cell frequency
$\beta=\sqrt5-2$. Eigenstates are computed with open
boundary conditions for $N_c=2^{13},\ldots,2^{17}$
two-site cells. Lyapunov estimates use transfer products
of up to $2\times10^6$ cells.

\begin{figure}[t]
    \centering
    \includegraphics[width=\linewidth]{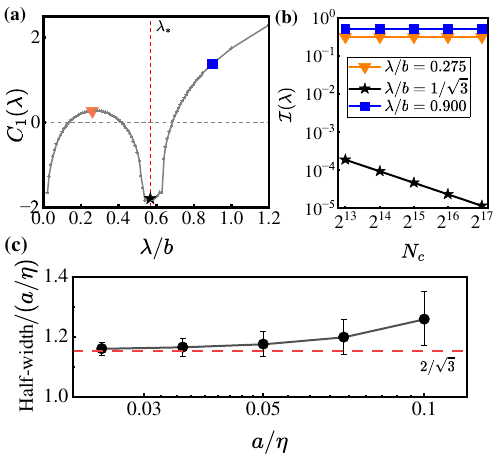}
    \caption{(a) Signed intercept $C_1(\lambda)$ and
    (b) cell IPR at the marked couplings for $a/b=10^{-3}$
    and $\eta=10^{-2}$. The IPR uses lower-band label $q=0.125$.
    The vertical dashed line marks $\lambda_*$. (c) Scaled recovery half-width along $a/b=\eta^{3/2}$.
    Error bars show its variation across 24 lower-band
    quantiles. The dashed line marks the first-order
    prediction $2/\sqrt3$.}
    \label{fig:3}
\end{figure}

\subsection{Recovery and its spectral extent}
\label{sec:spectralnumerics}

Figure~\ref{fig:3}(a) shows the sampled global intercept
$C_1(\lambda)$. Its sign follows the sequence negative,
positive, negative, and positive as $\lambda$ increases.
The positive intervals have positive exponents at all
sampled energies, while the recovered negative interval
contains the cancellation point $\lambda_*$.

To follow spectral populations, we use the quantile
$q\simeq j/(2N_c)$ of the ordered finite-chain eigenvalues,
recomputing its energy at each coupling and size.
For a normalized eigenstate, the cell probability and
cell inverse participation ratio are
\begin{equation}
 p_m=|x_m|^2+|y_m|^2,\qquad
 \mathcal I=\sum_m p_m^2.
 \label{eq:IPR}
\end{equation}
Summing within each cell removes the effect of shifting
weight between its two sites.

Figure~\ref{fig:3}(b) shows the IPR at the lower-band
quantile $q=0.125$, geometrically averaged over four
spectral phases. At $\lambda_*$, $\mathcal I$ decreases
approximately as $N_c^{-1}$ across the five sizes,
consistent with an extended state. At
$\lambda/b=0.275$ and $0.900$, it remains nearly
independent of size, consistent with localization.
The same spectral population therefore recovers spatial
extent between two localized regimes.

\begin{figure}[t]
    \centering
    \includegraphics[width=\linewidth]{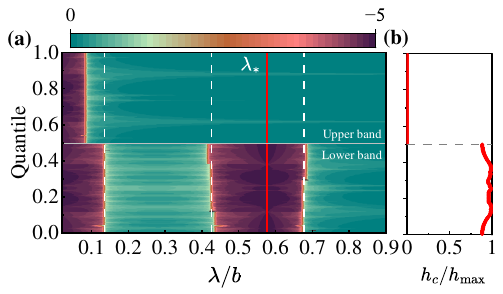}
    \caption{Recovery in the lower band for $a/b=10^{-3}$ and $\eta=10^{-2}$.
    (a) Cell IPR, shown as $\log_{10}\mathcal I$, versus coupling
    and spectral quantile $q$. White dashed lines mark the
    first-order $\rho=1$ crossings; the red line marks $\lambda_*$.
    (b) Normalized zero-exponent plateau width $h_c/h_{\max}$
    at $\lambda_*$, showing subcritical recovery in the lower band.
    The horizontal line at $q=1/2$ separates the two bands.}
    \label{fig:4}
\end{figure}

The participation map in Fig.~\ref{fig:4}(a)
shows that this recovery extends across the sampled
lower band. The upper band remains strongly concentrated
over the same coupling range. This band selectivity
agrees with the cancellation of the lower-state
modulation and the absence of a corresponding cancellation
in the upper state. The numerical boundaries coincide with the first-order $\rho=1$ crossings;
their median over quantiles differs from them by less than $4\times10^{-5}b$.

\begin{figure}[h]
    \centering
    \includegraphics[width=0.95\linewidth]{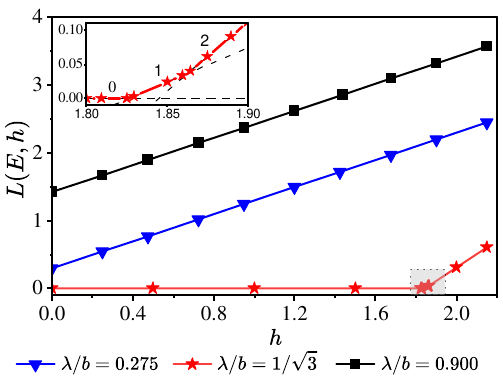}
    \caption{Complexified Lyapunov exponent $L(E,h)$ per cell
    at $q=0.125$ for the indicated couplings, with
    $a/b=10^{-3}$ and $\eta=10^{-2}$.
    The inset enlarges the recovery corner. Dashed lines show
    branch extrapolations, and numbers label their slopes.}
    \label{fig:5}
\end{figure}

\subsection{Lyapunov plateaus and recovery width}
\label{sec:complexnumerics}

Figure~\ref{fig:5} compares the complexified Lyapunov
exponent at $q=0.125$ for the same three couplings.
The spectral energies are selected at $\phi=0$, and
the profiles are averaged over three transfer phases.
Before and after recovery, $L(E,0)>0$. At $\lambda_*$,
the exponent is consistent with zero throughout a
finite interval of imaginary phase. The inset resolves
a short slope-one segment between the zero plateau
and the slope-two branch. This finite plateau identifies
the recovered state as subcritical.

Figure~\ref{fig:4}(b) extends this test across
the spectrum at $\lambda_*$. The plateau width
$h_c/h_{\max}$ is close to its allowed maximum throughout
the sampled lower band, while the upper band has no
resolved zero plateau. The recovered region therefore
combines spatial spreading with finite subcritical
plateaus.

Finally, Fig.~\ref{fig:3}(c) tests the predicted recovery
scale by varying $\eta$ along $a/b=\eta^{3/2}$.
For each $\eta$, we use $N_c=2^{17}$, $2\times10^6$
transfer cells, and three transfer phases. Recovery
edges are located separately at 24 lower-band quantiles
$q\in[0.02,0.48]$, using sign changes of $c_1$ from
Eq.~\eqref{eq:oneheight}. For each quantile, we locate the two couplings at which recovery begins
and ends, and take $w$ as half the distance between them. The bars show
how $w$ varies across the sampled quantiles. As $\eta$ decreases,
$w/(a/\eta)$ approaches $2/\sqrt{3}$ and its variation across the band
narrows. Extrapolating the three smallest values of $\eta$ to
$\eta\to0$ gives $1.154$, in agreement with $2/\sqrt{3}\simeq1.155$.
This supports the predicted $a/\eta$ recovery scale and its first-order
prefactor.

\section{Discussion}
\label{sec:discussion}

Increasing $\lambda$ strengthens the modulation
and shifts the weight between the two sites of each dimer.
This redistribution can reduce the potential seen by the
lower-band states even as the bare onsite amplitudes grow.
At first-order cancellation, the weak hopping remains
finite and the residual modulation is of order $\eta^2b$.
The hierarchy $\eta^2b\ll a\ll\eta b$ places the hopping
above this residual but below the modulation away from
cancellation, providing the scale separation for recovery.

The projected cancellation also persists under a change
of modulation ratio. Replacing the coefficients $3$ and
$1$ in Eq.~\eqref{eq:onsite} by fixed $s_A>s_B>0$ gives
\begin{equation}
 v_-^{(s)}(\lambda)=\frac{\eta\lambda}{2}
 \left[(s_A-s_B)-(s_A+s_B)
 \frac{\lambda}{\sqrt{b^2+\lambda^2}}\right],
 \label{eq:generalv}
\end{equation}
with projected cancellation point
\begin{equation}
 \lambda_*^{(s)}=
 \frac{b(s_A-s_B)}{2\sqrt{s_As_B}}.
 \label{eq:generalstar}
\end{equation}
This predicts how the cancellation shifts with the
modulation ratio.

Cancellation through redistribution of orbital weights
also appears in related molecular-band models
\cite{61}. Here, control of the residual terms and the exact transfer matrix connects
the projected mechanism to the full dimer chain. At the finite parameters
studied, the reentrant sign sequence of $C_1$, the spectrum-wide
participation map, the maximal zero-exponent plateaus across the lower
band, and the $a/\eta$ scaling of the recovery window support the same
picture. For $C_1<0$, the
quantity $-C_1$ gives the widest plateau width, retaining
information when the minimum physical exponent vanishes.

For sufficiently small nonzero $\eta$ under
Eq.~\eqref{eq:scaling}, the analytical result establishes a finite recovery interval in which every
spectral energy of a central window, holding one third of the lower-band
states, is subcritical. This interval is bracketed by uniform
positive-exponent regimes. The interval is sufficient
and does not identify exact critical couplings.
Positivity of the exponent alone is also not a universal
point localization theorem. The result provides
a controlled example of reentrance driven by changes
in the internal dimer weights during a fixed-profile
potential sweep.

\begin{acknowledgments}
This work was supported by the Scientific and Technological Council of
T\"urkiye (T\"UB{\.I}TAK) under Grant No.~125F435 and the Turkish Academy of
Sciences (TUBA) under Grant No.~AD-2026.  The numerical calculations reported
in this paper were performed at TUBITAK ULAKBIM, High Performance
and Grid Computing Center (TRUBA resources).
\end{acknowledgments}

\appendix

\section{Slope bound and large-phase asymptote}
\label{app:branches}

We take $a,b,\eta>0$ and $\lambda>0$. In Eq.~\eqref{eq:M},
the onsite factors at complex phase $z=\theta+ih$ are
\begin{align}
 P&=E-\lambda-3\eta\lambda\cos z,\nonumber\\
 Q&=E+\lambda+\eta\lambda\cos z.
\end{align}
Their product is
\begin{align}
 PQ&=E^2-\lambda^2
 -2\eta\lambda(E+2\lambda)\cos z\nonumber\\
 &\quad-3\eta^2\lambda^2\cos^2 z.
\end{align}
Thus every entry of $M_E$ has Fourier degree at most two.
Since $\cos(\theta+ih)=O(e^h)$ uniformly in $\theta$,
\begin{equation}
 L(E,h)\le\ln\sup_\theta\|M_E(\theta+ih)\|
 \le 2h+C,
\end{equation}
where $C$ is independent of $h$. Convexity makes the 
slopes of $L$ nondecreasing. A slope greater than two would
therefore contradict this bound. Evenness gives nonnegative
slopes for $h\ge0$, while acceleration quantization restricts
them to integers. Thus the allowed slopes are
zero, one, and two.

To determine the slope-two intercept, we use
\begin{equation}
 \cos(\theta+ih)
 =\frac12e^{h-i\theta}\left[1+O(e^{-2h})\right].
\end{equation}
The quadratic term in $PQ$ then gives
\begin{equation}
 M_E(\theta+ih)=
 -\frac{3\eta^2\lambda^2}{4ab}e^{2h-2i\theta}
 \begin{pmatrix}1&0\\0&0\end{pmatrix}
 +O(e^h),
\end{equation}
uniformly in $\theta$. After removing the scalar prefactor,
the matrices approach $\operatorname{diag}(1,0)$ with error
$O(e^{-h})$. For sufficiently large $h$, the remaining Lyapunov exponent is therefore $O(e^{-h})$, giving
\begin{equation}
 L(E,h)=2h+\ln\frac{3\eta^2\lambda^2}{4ab}
        +O(e^{-h}).
\end{equation}
This establishes Eq.~\eqref{eq:c2}. The intercept depends on
neither $E$ nor the irrational frequency $\beta$.

The slope-one intercept is defined by requiring the line $c+h$
to lie below $L(E,h)$ for every $h\ge0$. This requires
$c\le L(E,h)-h$ at every height, so the largest admissible
intercept is
\begin{equation}
 c_1(E,\lambda)=\min_{h\ge0}[L(E,h)-h].
\end{equation}
The large-$h$ asymptote gives
$L(E,h)-h=h+c_2(\lambda)+o(1)\to\infty$,
so the minimum is attained at a finite height. Consequently,
\begin{equation}
 L(E,h)\ge c_1(E,\lambda)+h,
\end{equation}
with equality at a minimizing height. If a slope-one segment
is present, $c_1$ is its intercept. The definition remains valid
when the supporting line touches only a corner or the endpoint
$h=0$, without an exposed slope-one segment. The slope bound makes
$L(E,h)-2h$ nonincreasing, with limit $c_2(\lambda)$, so
\begin{equation}
 L(E,h)\ge c_2(\lambda)+2h.
\end{equation}
At $h=0$, using $L(E,0)\ge0$ and the per-site normalization,
we obtain
\begin{align}
 \gamma(E,\lambda)
 &=\frac12L(E,0)\nonumber\\
 &\ge\max\left\{0,\frac12c_2(\lambda)\right\}\nonumber\\
 &=\max\left\{0,\ln\frac{\lambda}{\lambda_H}\right\},
 \qquad
 \lambda_H=\frac{2\sqrt{ab}}{\sqrt3\,\eta},
 \label{eq:physicalbound}
\end{align}

\section{Analytical bounds for the original chain}
\label{app:analytic}

\subsection{Recovery at the cancellation point}
\label{app:rotation}

We first show that the cancellation suppresses the modulation
in the exact transfer matrix. Set $r=a/b$,
$E=-2b/\sqrt3+ax$, and $c=\cos\theta$. At
$\lambda=\lambda_*=b/\sqrt3$, Eq.~\eqref{eq:M} gives
\begin{align}
 M_{11}&=-\frac{4x}{\sqrt3}+rx^2
 -\frac{2\eta x}{\sqrt3}c-\frac{\eta^2}{r}c^2,\nonumber\\
 M_{12}&=\frac1{\sqrt3}-rx-\frac{\eta}{\sqrt3}c,\nonumber\\
 M_{21}&=-\sqrt3+rx-\sqrt3\eta c,
 \qquad M_{22}=-r.
 \label{eq:exactMstar}
\end{align}

We restrict the scaled energy to
$I=[-\sqrt3/4,\sqrt3/4]$. To control the analytic dependence
on phase, we write $\theta=\vartheta+ih$, with real
$\vartheta$ and $h$, and consider the fixed strip $|h|<1$.
Throughout this domain,
\[
 |x|\le\frac12,
 \qquad
 |\cos\theta|\le\cosh1<2.
\]

Let $M_{\eta,x}$ denote the exact matrix in
Eq.~\eqref{eq:exactMstar} and $M_0(x)$ its 
limit in Eq.~\eqref{eq:M0}. Applying these bounds to the
entries of $M_{\eta,x}-M_0(x)$, and bounding the operator
norm by the sum of their absolute values, gives
\begin{equation}
 \|M_{\eta,x}-M_0(x)\|_{\rm strip}
 \le6\delta_\eta,\qquad
 \delta_\eta=r+\eta+\frac{\eta^2}{r}.
 \label{eq:stripbound}
\end{equation}
The strip norm denotes the supremum of the operator
norm over all real $\vartheta$ and $|h|<1$. The estimate holds
uniformly for every $x\in I$.

The relevant
smallness is set by $\delta_\eta$: choosing $r=\eta^p$ with
$1<p<2$ gives
\[
 \delta_\eta=\eta^p+\eta+\eta^{2-p}\longrightarrow0
 \qquad\text{as }\eta\to0.
\]
The exact transfer matrix therefore approaches the limiting
matrix throughout the chosen energy window.

The limiting propagation is oscillatory throughout $I$:
$|\operatorname{tr}M_0|\le1$, is inside the elliptic
regime $|\operatorname{tr}M_0|<2$. Each limiting matrix can
be transformed into a rotation by a change of
coordinates. Because the whole interval stays away from
$|\operatorname{tr}M_0|=2$, these transformations  remain uniformly bounded. In the same coordinates,
the exact matrix is a perturbation of a constant rotation
of order $\delta_\eta$.

\subsection{Subcriticality in the controlled window}
\label{app:subcritical}

The uniform matrix estimate constrains the entire spectral
window. Let $T_x$ be the change of basis
that transforms $M_0(x)$ into a rotation $R_x$. The transformed
exact matrix satisfies
\begin{equation}
 \widetilde M_{\eta,x}=T_x^{-1}M_{\eta,x}T_x,
 \qquad
 \|\widetilde M_{\eta,x}-R_x\|_{\rm strip}
 \le K_{\rm rot}\delta_\eta,
 \label{eq:rot}
\end{equation}
where $K_{\rm rot}$ is independent of $x\in I$. Since a rotation has operator
norm one, each transformed matrix has norm at most
$1+K_{\rm rot}\delta_\eta$. Multiplying these bounds and
using invariance of the Lyapunov exponent under the change of basis gives
\begin{equation}
 0\le L(E,h)\le
 \ln(1+K_{\rm rot}\delta_\eta),
 \qquad |h|<1.
 \label{eq:stripLEbound}
\end{equation}
The lower bound follows from the unit determinant.

Fix a height $0<H<1$. Since $\delta_\eta\to0$, we can
take $\eta$ sufficiently small that the upper bound in
Eq.~\eqref{eq:stripLEbound} is less than $H/2$.
In our phase normalization, every positive slope is at
least one. If such a slope appeared at
$t\le H/2$, convexity would give
\begin{equation}
 L(E,H)\ge L(E,t)+(H-t)\ge H/2,
\end{equation}
contradicting Eq.~\eqref{eq:stripLEbound}.
The exponent must therefore be constant for
$0\le h\le H/2$. Its evenness extends this plateau to
$|h|\le H/2$.

We now restrict to energies in the original spectrum.
If the plateau had positive height, the Lyapunov exponent would be
positive and would not change with $h$ near the real axis; such a cocycle
is called regular.
Avila's theorem would then imply uniform hyperbolicity. For nonzero hoppings $a$ and $b$, uniform
hyperbolicity of the cell transfer matrix is excluded
at spectral energies. Consequently,
\begin{align}
 L(E,h)&=0,\qquad |h|\le H/2,
 \quad
 E\in\sigma(H_{\lambda_*}),\nonumber\\
 E&=-2b/\sqrt3+ax,\quad x\in I.
 \label{eq:uniformplateau}
\end{align}
Every spectral energy in the controlled window is therefore subcritical.
We now show that the window carries a finite fraction of the spectrum.

In the coordinates of. (\ref{eq:rot}), each transfer step differs from the
rotation $R_x$ by at most $K_{\rm rot}\delta_\eta$. The angular advance per
cell therefore differs from that of $R_x$ by $O(\delta_\eta)$, uniformly in
$x\in I$, and the rotation number satisfies
\begin{equation}
  |\varrho(x)-\varrho_0(x)| \le B\delta_\eta ,
\end{equation}
where $\varrho_0(x)$ is the rotation number of $M_0(x)$. Since $M_0(x)$ has
unit determinant and $|\mathrm{tr}\,M_0|<2$, it is conjugate to a rotation
by the angle $2\pi\varrho_0$, with
\begin{equation}
  2\cos\bigl(2\pi\varrho_0(x)\bigr) = \mathrm{tr}\,M_0(x) = -\frac{4x}{\sqrt{3}} .
\end{equation}
At the window edges $x=\mp\sqrt{3}/4$ the trace equals $\pm1$, giving
$2\pi|\varrho_0| = \pi/3$ and $2\pi/3$. The negative lower-left entry of
$M_0$ fixes the direction, so $\varrho_0$ covers
$[-1/3,-1/6]$ across $I$.

The cell matrix $M_E$ is the product of the two one-site transfer matrices,
$(x_m,y_{m-1})\to(y_m,x_m)\to(x_{m+1},y_m)$. As in oscillation theory, each
state below $E$ adds a half-turn to the solution at energy $E$, so the
rotation number determines the integrated density of states~\cite{64,65}.
Across the full lower band, $|x|\le\sqrt{3}/2$, the trace runs from $2$ to
$-2$ and $|\varrho_0|$ covers $[0,1/2]$. The window covers an interval of
length $1/6$ within it, and therefore contains a fraction
$1/3+O(\delta_\eta)$ of the lower-band states. Since the integrated density
of states increases only on the spectrum, these states are in the spectrum,
and by Eq.~(\ref{eq:uniformplateau}) all of them are subcritical.

\subsection{Finite recovery interval}
\label{app:interval}

We now show that recovery survives a small change in
$\lambda$, rather than occurring only at the cancellation
point. Write $\Delta\lambda=\lambda-\lambda_*$ and follow
the lower-cell level by setting
\[
 E=-\sqrt{b^2+\lambda^2}+ax,
 \qquad x\in I.
\]
The exact transfer entries are
\begin{align}
 M_{11}={}&-\frac{2\sqrt{b^2+\lambda^2}}{b}x+rx^2
 \nonumber\\
 &-\frac{2\eta\lambda}{ab}
 \left[2\lambda-\sqrt{b^2+\lambda^2}+ax\right]\cos\theta
 \nonumber\\
 &-\frac{3\eta^2\lambda^2}{ab}\cos^2\theta,
 \nonumber\\
 M_{12}={}&\frac{\sqrt{b^2+\lambda^2}-\lambda-ax
 -\eta\lambda\cos\theta}{b},\nonumber\\
 M_{21}={}&\frac{-\sqrt{b^2+\lambda^2}-\lambda+ax
 -3\eta\lambda\cos\theta}{b},\nonumber\\
 M_{22}={}&-r.
 \label{eq:generalM}
\end{align}

The cancellation is controlled by
$2\lambda-\sqrt{b^2+\lambda^2}$. This expression vanishes
at $\lambda_*=b/\sqrt3$, and its derivative there is $3/2$.
For $|\Delta\lambda|\ll b$,
\begin{equation}
 2\lambda-\sqrt{b^2+\lambda^2}
 =\frac32\Delta\lambda
 +O\!\left(\frac{\Delta\lambda^2}{b}\right).
 \label{eq:detuningexpansion}
\end{equation}
Its contribution to $M_{11}$ is therefore of order
$\eta|\Delta\lambda|/a$. This is the restored first-order
modulation measured relative to the hopping.

The phase-independent terms change by order
$|\Delta\lambda|/b$. The terms proportional to $ax\cos\theta$
and $\cos^2\theta$ remain of order $\eta$ and $\eta^2/r$,
respectively, while the weak-hopping corrections remain
of order $r$. Using the same bounds on $x$ and $\cos\theta$
as at cancellation gives
\begin{equation}
 \|M_E-M_0(x)\|_{\rm strip}
 \le C\left[
 \delta_\eta+\frac{|\Delta\lambda|}{b}
 +\frac{\eta|\Delta\lambda|}{a}
 \right],
 \label{eq:intervalbound}
\end{equation}
for $x\in I$ and $\lambda$ in a fixed neighborhood
of $\lambda_*$. The constant $C$ is independent of
$\eta$ and $\Delta\lambda$ in this neighborhood.

To keep this error small, choose
$|\Delta\lambda|<c_*a/\eta$, with fixed $c_*>0$.
Since $a/b=\eta^p$, Eq.~\eqref{eq:intervalbound} becomes
\begin{equation}
 \|M_E-M_0(x)\|_{\rm strip}
 \le C\left[
 \delta_\eta+c_*\eta^{p-1}+c_*
 \right].
 \label{eq:windowerror}
\end{equation}
First choose $c_*$ so that $Cc_*$ is below the smallness
threshold required by the recovery argument. Then take
$\eta$ sufficiently small: both $\delta_\eta$ and
$\eta^{p-1}$ vanish because $1<p<2$. The same branch argument applies throughout this interval.
Equation~\eqref{eq:intervalbound} keeps the transformed
matrix uniformly close to a rotation when $c_*$ is chosen
sufficiently small and then $\eta$ is reduced. Hence every
spectral energy in the window
$E=-\sqrt{b^2+\lambda^2}+ax$, $x\in I$, is subcritical.
The rotation-number estimate of Appendix \ref{app:subcritical} applies with the same
uniformity, so at every coupling in this interval the window holds a
fraction $1/3+O(\delta_\eta)$ of the lower-band states, all subcritical.
This establishes Eq. (\ref{eq:rigorouswindow}).

The first-order AA description gives a separate leading
estimate of the interval width. Its projected modulation is
\[
 v_-(\lambda)=\eta\lambda
 \left(1-\frac{2\lambda}{\sqrt{b^2+\lambda^2}}\right).
\]
Using Eq.~\eqref{eq:detuningexpansion},
\begin{equation}
 v_-(\lambda)
 =-\frac{3\eta}{4}\Delta\lambda
 +O\!\left(\frac{\eta\Delta\lambda^2}{b}\right).
 \label{eq:projecteddetuning}
\end{equation}
The leading hopping at cancellation is
$|t_{\rm eff}(\lambda_*)|=\sqrt3a/4$. Linearizing the
modulation and evaluating the hopping at this point gives
\[
 \rho(\lambda)=\frac{|v_-(\lambda)|}{2|t_{\rm eff}(\lambda)|}
 \simeq\frac{\sqrt3\eta}{2a}|\Delta\lambda|.
\]
On the scale $|\Delta\lambda|\sim a/\eta$,
$|\Delta\lambda|/b\sim\eta^{p-1}\to0$.
The linearization and the use of the hopping at $\lambda_*$
are therefore consistent to leading order.
The AA condition $\rho<1$ gives the estimated width
\[
 |\lambda-\lambda_*|\lesssim\frac{2a}{\sqrt3\,\eta},
\]
as quoted in Eq.~\eqref{eq:projectedwindow}.

\subsection{Positive exponent before recovery}
\label{app:early}

We show that every spectral energy has a positive physical
exponent at the earlier coupling $\lambda_{\rm e}=b/4$.
Assume $\eta\le10^{-4}$ and $r=a/b\le\eta/4$.
The weak hopping and modulation shift the spectrum by at
most $\eta b$ from the isolated-cell energies
$\pm\sqrt{17}b/4$. Consequently,
\begin{align}
 |E^2-b^2-\lambda_{\rm e}^2|
 &\le\frac94\eta b^2,\nonumber\\
 |E+\lambda_{\rm e}|&\ge\frac{3b}{4},
 \qquad
 |E+2\lambda_{\rm e}|\ge\frac b2.
 \label{eq:spectralbounds}
\end{align}
These bounds hold throughout both spectral bands.

Eliminating the $B$ amplitudes gives an exact equation
for the $A$ amplitudes $x_m$:
\[
 x_{m+1}=Q_E(\theta_m)x_m-R_E(\theta_m)x_{m-1}.
\]
Define
\[
 d_E(\theta)=E+\lambda_{\rm e}
             +\eta\lambda_{\rm e}\cos\theta.
\]
The coefficients are
\begin{align}
 R_E(\theta)
 &=\frac{d_E(\theta)}{d_E(\theta-2\pi\beta)},\nonumber\\
 Q_E(\theta)
 &=\frac{1}{ab}\Big[
 E^2-b^2-\lambda_{\rm e}^2
 -2\eta\lambda_{\rm e}(E+2\lambda_{\rm e})\cos\theta
 \nonumber\\
 &\hspace{33mm}
 -3\eta^2\lambda_{\rm e}^2\cos^2\theta
 \Big]-rR_E(\theta).
 \label{eq:eliminated}
\end{align}

We first establish growth at the imaginary phase
$h=\ln100$. This shift enhance the cosine:
$|\cos(\theta+ih)|\ge\sinh h>49$.
The term linear in $\cos\theta$ therefore dominates the
phase-independent term, while the extra factor of $\eta$
keeps the quadratic correction small.
The denominator also stays away from zero:
$b/2\le|d_E(\theta+ih)|\le3b/2$.
Using Eq.~\eqref{eq:spectralbounds}, we obtain
\[
 |R_E(\theta+ih)|\le3,
 \qquad
 |Q_E(\theta+ih)|\ge\frac{8\eta}{r}.
\]
Thus the coefficient multiplying the current amplitude
is much larger than the coefficient multiplying the
previous one.

To see how this produces sustained growth, start with
a transfer vector satisfying $|v_2|\le|v_1|$.
The next vector is
$(v'_1,v'_2)=(Q_Ev_1-R_Ev_2,v_1)$, so
\[
 |v'_1|
 \ge\bigl(|Q_E|-|R_E|\bigr)|v_1|
 \ge\frac{7\eta}{r}|v_1|,
 \qquad
 |v'_2|\le|v'_1|.
\]
The same condition therefore holds after every step,
and the amplitude grows by at least $7\eta/r$ per cell.
Eliminating the $B$ amplitudes is a bounded change of
transfer coordinates and does not change the Lyapunov
exponent. Hence
\[
 L(E,h)\ge\ln\frac{7\eta}{r}.
\]

The maximum slope of $L(E,h)$ is two. Returning from
$h=\ln100$ to the real axis can therefore reduce the
exponent by at most $2\ln100$. With the per-site
normalization $\gamma=L/2$, this gives
\begin{equation}
 \gamma(E,\lambda_{\rm e})
 \ge\frac12\ln\frac{7\eta}{10^4r}.
 \label{eq:explicitbound}
\end{equation}
For $r=\eta^p$ with $p>1$, the ratio $\eta/r$ diverges
as $\eta\to0$. The bound is therefore positive at every
spectral energy for sufficiently small $\eta$.

Together with the recovered subcritical spectral set
and the later positive-exponent bound in
Eq.~\eqref{eq:physicalbound}, this establishes recovery
between two regimes with uniformly positive exponents.

\bibliography{bib}
\end{document}